\def\apjs{\ {ApJS}\ }

\def\pasj{\ {Publ. Astr. Soc. Japan}\ }

\def\prd{\ {Phys. Rev. D}\ }

\documentclass[]{elsart}
\usepackage{epsfig}

\usepackage{graphicx}
\usepackage{psfig}

\usepackage{ifthen}
\newboolean{SimonsSetup}
\setboolean{SimonsSetup}{true}

\def\apgt{\ {\raise-.5ex\hbox{$\buildrel>\over\sim$}}\ } 
\def\aplt{\ {\raise-.5ex\hbox{$\buildrel<\over\sim$}}\ } 
\def\lt{\ {\raise-.5ex\hbox{$\buildrel>$}}\ } 
\def\gt{\ {\raise-.5ex\hbox{$\buildrel<$}}\ }

\begin{document}
\runauthor{Portegies Zwart at al.}
\begin{frontmatter}

\title{Simulating the universe on an intercontinental grid of supercomputers}

\author{Simon Portegies Zwart$^{1}$,} 
\author{Tomoaki Ishiyama$^{2}$,} 
\author{Derek Groen$^{1,3,4}$,}
\author{Keigo Nitadori$^{5}$,}
\author{Junichiro Makino$^{6}$,} 
\author{Cees de Laat$^{7}$,} 
\author{Stephen McMillan$^{8}$,}
\author{Kei Hiraki$^{9}$,} 
\author{Stefan Harfst$^{1}$,}
\author{Paola Grosso$^{7}$,} 

{\noindent 
$^{1}$ Sterrewacht Leiden, P.O. Box 9513, 2300 RA Leiden, The Netherlands\\
$^{2}$ Department of General and Sciences, University of Tokyo, Tokyo 153-8902, Japan \\
$^{3}$ Astronomical Institute "Anton Pannekoek" , University of Amsterdam, Amsterdam, The Netherlands\\
$^{4}$ Section Computational Science, University of Amsterdam, Amsterdam, The
Netherlands\\
$^{5}$ Department of Astronomy, School of Science,
University of Tokyo, Tokyo 113-8654, Japan\\
$^{6}$ Center for Computational Astrophysics in Tokyo, Japan\\
$^{7}$ Section System and Network Engineering Science, University of
Amsterdam, Amsterdam.\\
$^{8}$ Department of Physics, Drexel University, Philadelphia, PA
       19104, USA; \\
$^{9}$ Department of Creative Informatics,
Graduate School of Information Science and Technology
the University of Tokyo.\\
}

\begin{abstract}

Understanding the universe is hampered by the elusiveness of its most
common constituent, cold dark matter. Almost impossible to observe,
dark matter can be studied effectively by means of simulation and
there is probably no other research field where simulation has led to
so much progress in the last decade.  Cosmological N-body simulations
are an essential tool for evolving density perturbations in the
nonlinear regime.  Simulating the formation of large-scale structures
in the universe, however, is still a challenge due to the enormous
dynamic range in spatial and temporal coordinates, and due to the
enormous computer resources required.
The dynamic range is generally dealt with by the hybridization of
numerical techniques.  We deal with the computational requirements by
connecting two supercomputers via an optical network and make them
operate as a single machine. This is challenging, if only for the fact
that the supercomputers of our choice are separated by half the
planet, as one is located in Amsterdam and the other is in Tokyo.  The
co-scheduling of the two computers and the 'gridification' of the code
enables us to achieve a 90\% efficiency for this distributed
intercontinental supercomputer.  We conclude that running cosmological
$N$-body simulations on a limited number of ($\aplt 100$) processors
concurrently on more than 10 supercomputers in ring topology with a
high-bandwidth network would provide satisfactory performance and is
politicaly favorable regarding the acquisition of the resources.

\end{abstract}
\end{frontmatter}

Computing Methodologies: Simulation, Modeling, and Visualization: Distributed

\section{Introduction}

Since the beginning the size and complexity of the universe have been
increasing continuously.  As a consequence we observe today large
structures consisting of galaxies.  The visible superstructure of the
universe is made of baryonic material, e.g.\, stars and gas. But the
majority of the mass is in the form of dark matter, which is affected
by gravity but does not interact electro-magnetically.  The best model
for the formation and evolution of this superstructure is called
$\Lambda$ cold dark-matter cosmology
($\Lambda$CDM)\cite{1981PhRvD..23..347G}, according to which the
universe is about 13.7 billion years old and comprises of about 4\%
baryonic matter, $\sim 23$\% non-baryonic (dark) matter and 72\%
(dark) energy, indicated by the letter
$\Lambda$ \cite{2007ApJS..170..377S}.
The nature of dark matter is unknown and from an observational
perspective it is hard to resolve this lacuna in our understanding.
Evidence for dark matter comes from a variety of
observations, which includes the 
rapid rotation of the Milky-Way and other galaxies: its stars would be
flung out in the absence of dark matter.

Weakly-interacting massive particles form the most promising
explanation for dark matter
\cite{1977PhRvL..39..165L}, in which case the entire
universe is filled with a finely grained substance that behaves like a
gravitational fluid but is invisible otherwise.  The gravitational
force in the Newtonian limit $\propto 1/r^2$, which is a long-range
force in particular since the enclosed mass scale $\propto
r^3$. Together with the absence of a shielding mechanism, even very
distant objects cannot be ignored.  The way in which dark matter
interacts is therefore rather simple and well understood. As a result,
we can effectively study dark matter by simulation and use these
results to understand observations and to make predictions.

One of the most favorable techniques to study the formation of large
scale structure in the universe is by means of gravitational $N$-body
simulations, in which each particle in the simulation represents the
fluid of dark matter particles. 
The most widely used algorithm is the treePM method, in which the
short range forces are resolved using a Barnes-Hut tree code, whereas
the long range interactions are simulated using a Particle-Mesh
method \cite{1995ApJS...98..355X,Hockney1988}. This combination of
methodologies provide good performance compared to using only a tree
code but still at a reasonable accuracy compared to a pure
particle-mesh technique, as in the treePM method we resolve short as
well as long range forces reasonably well.  In addition, the algorithm
scales well to a large number of processors by optimized domain
decomposition (Ishiyama et al. in press).
Since simulating the entire
universe is not really possible (yet) we study a small part, using
periodic boundary conditions to mimic {\em ad infinitum}.

The computational demand for our large scale cosmological $N$-body
simulation is enormous, and instead of running on a single
supercomputer, we opted for running concurrently on two widely
separated supercomputers. We demonstrate that that it can be efficient
to run high-performance production simulations on a computational
grid.

\section{The intercontinental grid}

The future of large-scale scientific computing infrastructures aims at
distributing resources rather than concentrating supercomputers
locally \cite{2008PCAA.book.....H}. The higher cost effectiveness of
distributing resources can be efficient for those applications in
which the compute time scales steeper than the communication, or when
the problem can be decomposed in domains \cite{FosterKessleman}.

Instead of starting the simulation on one location and switch half way
to another computer to continue the calculation, we run the simulation
on two supercomputers concurrently.  One of our computers is located
in Amsterdam (the Netherlands) and the other is in Tokyo (Japan).  The
challenge is to efficiently use a large number of processors separated
by half the planet without running in the inter-processor and
intercontinental communication bottlenecks. If we can run successfully
among two supercomputers separated by half the planet, we can be
confident about upscaling our virtual organization to include more
than two supercomputers.

The management, political and technical issues of coupling the
computers with an uncongested network and to schedule the resources
proves to be extremely challenging, and one can wonder if it is worth
the effort. The preparations for our calculations lasted about a year.
However, running on a single supercomputer would have required its
entire capacity for several months, which would probably not have been
granted.  Acquiring relatively little time on a large number of
supercomputers proves to be considerably easier. With the expertise
obtained in running on two supercomputers we can now extend to more
sites without much additional overhead.  Of course, the political and
technical issues of coupling the computers with an efficient network
and to schedule the resources remain challenging, but many of these
aspects will become easier once high-performance grid computing
becomes mainstream.  Additional complications arise by the required
hybrid parallelization strategies, the diversity in topologies,
scheduling, load balancing and the complications introduced by the
different hardware architectures in particular if, as in our case, the
code is machine dependent.

\begin{figure}[h] \ifthenelse{\boolean{SimonsSetup}}{ \psfig{figure=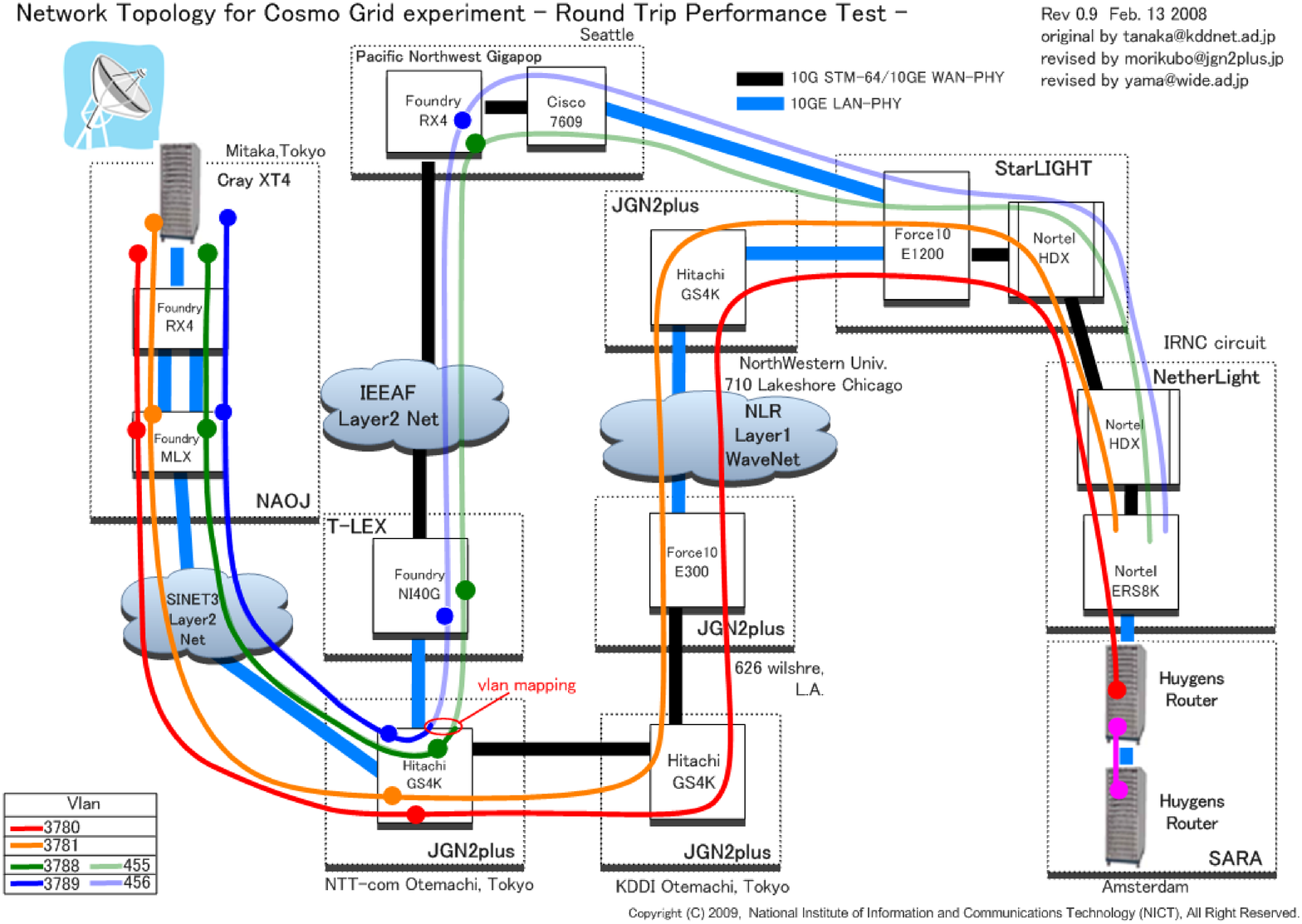,width=\columnwidth}
     }{} \caption[]{
     Network topology for the ComoGrid experimental setup. We used two
     main network connections between JGN2plus and StarLIGHT; one of
     then goes via L-LEX and Gigapop, the other connects directly from
     StarLIGHT to JGN2plus.
     Both paths have a network bandwidth of 10Gb/s, carried over
     10~GE-LANPHY and 10~GE-WANPHY segments depending on the type of
     interfaces present at the various switching points. Thick blue
     lines (LANPHY) and thick black lines (WANPHY).
  \label{Fig:Network}
  }
  \end{figure}

One of the complications we encountered is related to the network
topology and internal hardware setup.  Neither supercomputer is
directly connected to the optical network, but communication is
realized via a special node that is connected to the outside world
with a 10GbE optical switch in Amsterdam and Neterion NIC in
Tokyo. Upon every communication step each processor: identifies the
particles which need to be communicated, packs them and sends the
particles to the supercomputer's internal communication node, which
subsequently sends the entire package of particles to the
communication node outside the firewall. This package is subsequently
transmitted to the distant computer 9,400\,km away, as the bird
flies. The data however, travels $\sim 27,000$\,km as the network
links criss-crosses the Atlantic ocean, the USA and the Pacific
ocean. In Fig.\,\ref{Fig:Network} we illustrate the network topology.
With the speed of light in fiber this distance is covered in 0.138
seconds, resulting in a round-trip time of 0.277 seconds.

The network communication was realized by configuring two virtual
LANs~(vLAN) to create two paths connecting both supercomputers. One
vLAN was used for the production traffic and runtime synchronization,
while the second was configured for testing purposes and for
collecting the simulation data at the PByte storage array in
Amsterdam.  Both paths are dedicated to our experiments which prevents
packet loss, packet reordering and latency changes due to congestion.
In the secondary path we had to execute vLANs translation in the JNG
network, because the chosen vLANs numbers were unavailable along the
entire path length. The choice of an available vLAN number for the
end-to-end communication should have been trivial; but since our
multi-domain light path lagged automatic configuration tools, human
intervention was necessary.  The time required for solving problems
depends on the quick responses to email and phone calls, which is
hindered by working over several time zones; debugging link failure at
Layer2 in a multi-domain multi-vendor infrastructure is impractical
and extremely time consuming.  A self-healing or automatic setup would
enormously help future experiments.

The Research and Education Networks (REN) provided the links for free,
which is motivated by their vision to advertise and broaden their
services to the scientific community.  In the last few years many RENs
have adopted a hybrid model for their architecture, expanding it to
routed IP services where users have access to dedicated light paths in
which communication occurs at lower layers of the open
system-interconnection reference model. We settled for a flat Ethernet
path, but still had to make the network reservations 3 weeks in
advance of each simulation restart, and then it would generally take
about a week before it was fully operational.  Our overall experience
with the use of light paths however, is quite positive. We could not
have performed our calculation without this type of network setup and
certainly perceived the advantage of the unrestricted and exclusive
use of the links.

\subsection{Demand on the network}

While performing a cosmological $N$-body simulation in which the
computational domain is shared by two (or more) computers, the
positions and velocities of all particles on both machines have to be
synchronized throughout the simulation. This introduces an enormous
demand on the network between the computers.  Luckily
only a small boundary layer between the two halves of the universe and
the layer nearest the periodic boundary are communicated between
Amsterdam and Tokyo.  The amount of data that has to be communicated
per step is then
\begin{equation}
  S_{\rm comm} = 144 N^{2/3} + 4N_p^3  + 4N/S_r\,\,\, {\rm Byte}.
\label{Eq:CommSize}
\end{equation}
Here $N$ is the number of particles, $N_p$ is the resolution of the
mesh in one dimension and $S_r$ is the sampling rate, which for the
production simulation $S_r = 5000$.  The first term is required for
communicating the tree structure, the second term is for exchanging
the particles in the border region, and the last term is for
exchanging the mesh.  The first term in this estimate depends slightly
on redshift ($z$) and on the opening angle in the treecode ($\theta$),
but is accurate for $\theta = 0.5$.


%
%
%
%

While running, dense dark-matter clumps may not be distributed evenly
across the computational domain and load balancing is done by
guaranteeing that the calculation time per step is the same on each
computer, generating a variable boundary layer between the two
computers. Where initially both computer resolve exactly half the
universe, in due time one of the two computers tends to deal with a
larger volume.

The communication within each of the supercomputers is realized by
domain decomposition, using the Message Passing Interface ({\tt
mpich} \cite{Gropp:1996:HPI}). To warrant efficient and stable data
transfer we developed the parallel socket library {\tt MPWide} to
facilitate the communication outside the local MPI domains (Groen etal
2009, in preparation). This communication consists of data transfers
between the local compute nodes and the local communication node, and
for the data transfer over the light-path between the supercomputers.

The socket library is included in all processes, providing an
interface similar to regular MPI. A static communication topology is
established at startup which uses multiple {\tt tcp} connection for
each intra-cluster communication path and multiple {\tt tcp}
connections for paths between supercomputers.  The concurrent use of
multiple streams is realized by running a separate thread for each
stream.  For the smaller runs we used 16 {\tt tcp} streams, and 64
streams for the larger runs (see Tab.\,\ref{Tab:Simulation}).

\section{The Simulation environment}

We adopted the treePM code {\tt GreeM} which was initially developed
by \cite{2005PASJ...57..849Y} and rewritten by Ishiyama etal (in
press) to run efficiently on the selected hardware. The equations of
motion are integrated in co-moving coordinates using the leap-frog
scheme with a shared but variable time-step.
Good performance at sufficient accuracy is then achieved when the
time step is at least an order of magnitude smaller than the crossing
time in the densest halo \cite{Hockney1988}.  Spurious relaxation
effects in the high-density clumps are prevented by introducing a
Plummer softening to the force, which is comparable to the local
inter-particle distance.

Since most simulation time is spent in calculating the gravitational
forces between particles, we optimize this operation using the single
precision X86-64 Streaming SIMD Extensions (SSE).  The inverse
square-root SSE instruction provides the best speedup in calculating
Newtonian gravity.  We further improve performance by minimizing RAM
access through the 16~XMM registers for the force operations, and by
operating on pairs of two 64-bit floating point words concurrently
(Nitadori \& Yoshikawa in preparation).

With these optimizations the Power6 with symmetric multiprocessing is
per core is about 4.2\% slower than the Intel based Cray XT4. This
small difference in speed is achieved by adopting the x86 SSE version
and the Power AltiVec architectures, but for the former we use in-line
assembly whereas for the latter we adopted the intrinsic functions
(Nitadori \& Yoshikawa in preparation).

We measure the performance of the code using the Amsterdam and Tokyo
machines separately with $p=1$ to $p=1024$ processors and $N=256^3$ to
$N=1024^3$ particles.  The wall clock time $t_{\rm cpu}$ is then
fitted by $t_{\rm cpu} \simeq 14,500p^{-0.91}$ seconds, with a ten
fold increase of $t_{\rm cpu}$ for every increase of the number of
particles by $2^3$.  For relatively small $N$ we lose scalability with
respect to the number of processors for $p \simeq 10^3$ and $N \aplt
256^3$.

\section{Simulating the Universe}

We are interested in structures of kpc to Mpc size.  To minimize the
effect of the periodic boundary conditions on such dark matter
distributions we selected our simulation box at
$z=0$\footnote{Redshift is the standard measure of time in cosmology:
the universe was born $z\rightarrow\infty$ and today $z\equiv0$.}  to
have sides of 30\,Mpc. The initial dark-matter distribution is
generated at $z \simeq 65$, assuming that the relation between the
velocity and the potential in Zel'dovich approximation is the same as
in the linear theory \cite{2007PhRvD..76j3505J}. The density field is
then realized by multi-scale Gaussian random fields, which is
described in terms of its power spectrum and was generated using {\tt
MPGRAFIC} \cite{2008ApJS..178..179P}.

We further adopted cosmological parameters which are consistent the
5-year WMAP results \cite{2008arXiv0803.0547K}.  For clarity we opted
for the nearest round values, which yields: matter (including dark)
density $\Omega_m = 0.3$, dark energy density $\Omega_\Lambda = 0.7$,
the slope for the scalar perturbation spectrum $n_s = 1.0$, and the
amplitude of fluctuations $\sigma_8 = 0.8$. With these parameters the
universe is about 13.7 billion years old and the Hubble constant
$H_0 \simeq 70$\,km/s/Mpc.

We ran several realizations at different resolution (in mass as well
as spatially) to measure the performance before we completed a
simulation to $z=0$. An overview of the performed simulations is
presented in Tab.\,\ref{Tab:Simulation}.
  
   \begin{table}[htbp!]
    \begin{tabular}{rrcrrrr}
    \hline
      N    & $N_p$& p (A+T) &$t_{\rm tot}$&$t_{\rm cpu}$ &$\eta_p$\\
   16777216& 128  &  30+30  &  31.2 &  23.58 & 1.51 \\
  134217728& 256  &  30+30  & 116.8 & 105.7  & 1.81 \\
  134217728& 256  &  60+60  &  71.9 &  58.8  & 1.64 \\
  134217728& 256  & 120+120 &  53.0 &  36.8  & 1.39 \\
 1073741824& 256  & 500+250 &  60.0 &  40.0  & ---  \\
 8589934592& 256  & 500+250 & 380.0 & 310.0  & --- \\
     \hline
     \end{tabular}
      \caption{The performed simulations in a computational box of
      30\,Mpc starting at $z\simeq 65$, a softening of 300\,pc and the
      opening angle in the tree-code $\theta = 0.3$ for $z>10$ after
      which $\theta = 0.5$.  The first column gives the number of
      particles in the simulation followed by the number of mesh-cells
      in one dimension. Then we give the number of processors in
      Amsterdam (A) and Tokyo (T). The next two columns give the
      average wall-clock time for one step ($t_{\rm tot}$) and the
      time spent computing the forces ($t_{\rm cpu}$), both in
      seconds. The last two columns gives the speed-up $\eta_p$ which
      is defined as the wall-clock time of a single computer as
      fraction of the wall-clock time of the grid of supercomputers.
      Here we define $\eta_p$ for a grid where each computer has the
      same number of processors $p$. For the last two entries $\eta_p$
      is then not properly defined.
%
        \label{Tab:Simulation}
        } 
     \end{table}

\begin{figure}[h] \ifthenelse{\boolean{SimonsSetup}}{ \psfig{figure=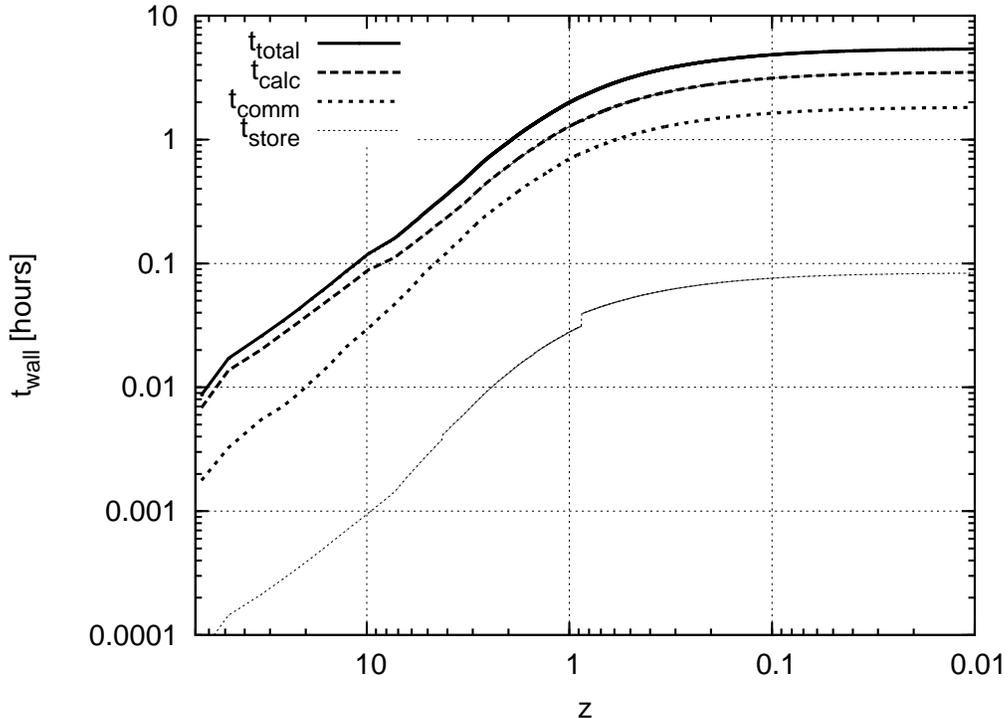,width=1.0\columnwidth}
  }{}
   \caption[]{Wall-clock time as a function of redshift ($z$) for the
   simulation with $N=16777216$.  The solid curve gives the wall-clock
   time, including the network wait stages, the dashed curve gives
   calculation time and the dotted curve indicates the time spend
   communicating. The lower thin dotted curve gives the time for
   storing data.}
   \label{fig:CPUtimevsZ} 
   \end{figure}

The simulation with $N=16777216$ took about 10 hours to complete from
$z \simeq 65$ to $z=0$.  In Fig.\,\ref{fig:CPUtimevsZ} we show the wall-clock
time of this run decomposed in the time spend in calculation,
communication and storing the data to the file system. About 25\% of
the time is spent in communication (see Tab.\,\ref{Tab:Simulation}),
which is roughly constant through the simulation.  We encountered a
few problems between $z\simeq16$ and $z\simeq15$, which are probably
related to packet loss in the network transport protocol suite and
resulted in an additional performance loss of a few per-cent.

During the communication the speed of data transfer averages
21.1Mbit/s with peaks of about 7Gbit/s and per step between 16MB and
26MB is transferred.  For the larger runs the average throughput
increases to about 208Mbit/s since the size of the packages increases
with $N$ (see Eq.\,\ref{Eq:CommSize}).

\begin{figure}[ht]
  \ifthenelse{\boolean{SimonsSetup}}{
   \psfig{figure=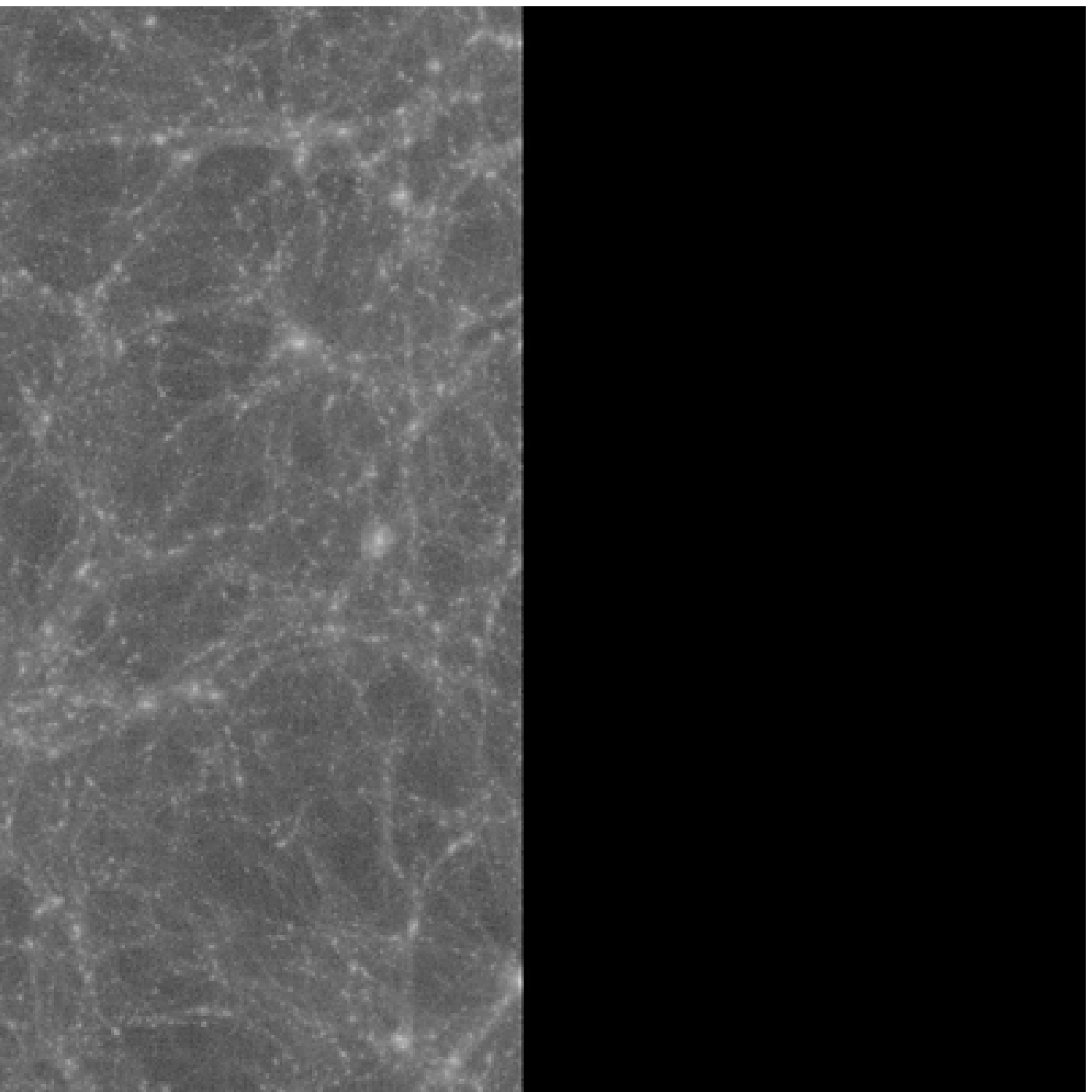,width=0.45\columnwidth}
   ~\psfig{figure=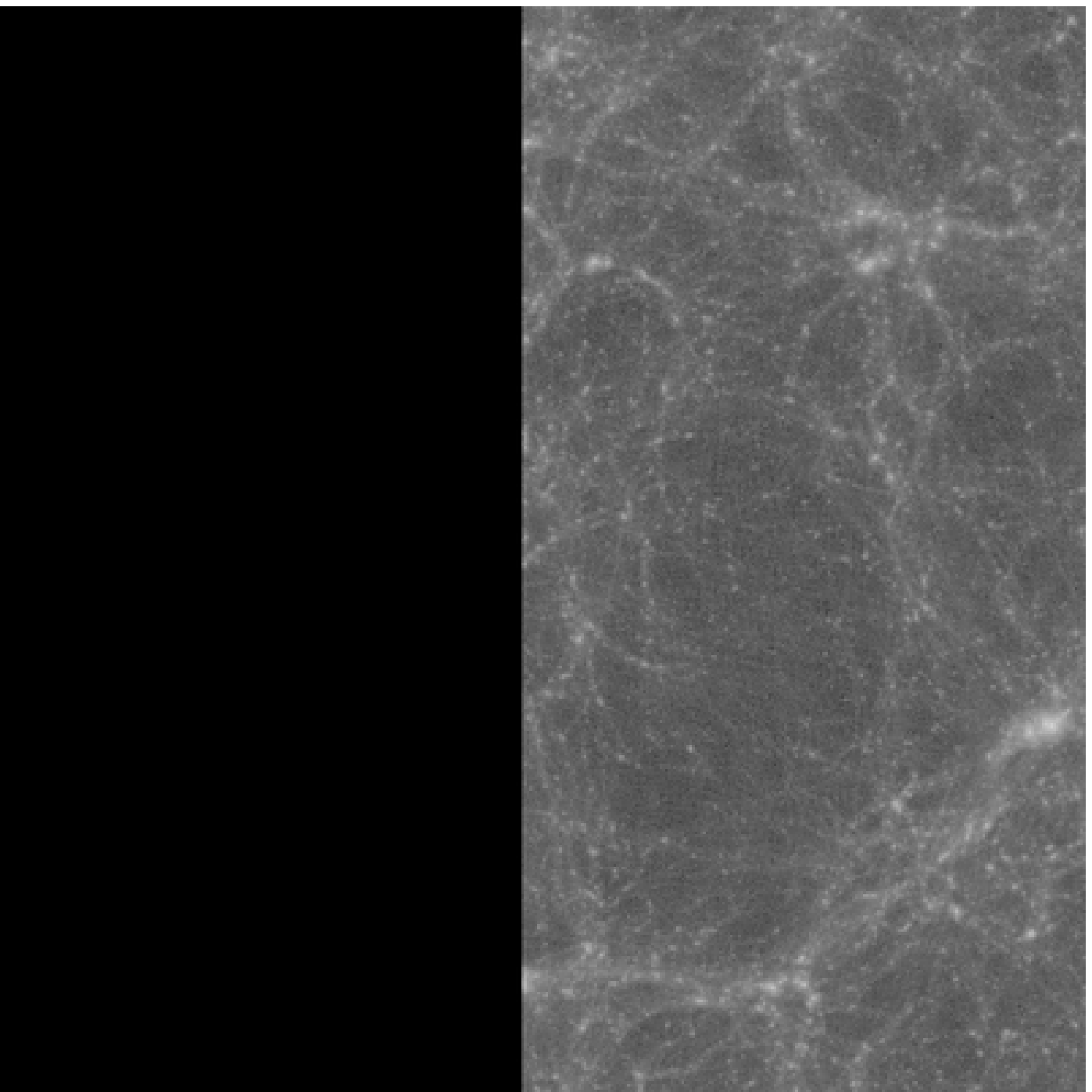,width=0.45\columnwidth}
  }{}
   \caption[]{Final snapshot of the simulation with $N=256^3$ with
              Tokyo (left) and Amsterdam (right) .  The black part of
              each figure represents the memory share that is located
              at the other site.  }
   \label{fig:CGN256qToZ0} 
   \end{figure}

The result of the simulation with $N=16777216$ at $z=0$ is presented
in two panels in Fig.\,\ref{fig:CGN256qToZ0}.  Each of the two panels
show the universe as it is seen by the two supercomputers, with the
black parts on the right of the left panel (left on the right image)
indicate the part of the universe that resides on the other machine.

\section{Concluding remarks}

The formation of large scale structure in the universe can be studied
effectively by means of simulation but requires phenomenal computer
power. Even with fully optimized numerical methods and approximations
the wall-clock time for such simulations can exceed several months.
Our largest simulation with $N=8589934592$ has been running for about
1 million CPU hours (more than 1 month on 1024 processors) to reach
$z\simeq 1.5$ and we expect to spend another $\sim 2$~million CPU
hours to reach $z=0$.

The runs in Tab.\,\ref{Tab:Simulation} are performed on a grid of
supercomputers. With our implementation the latency and throughput of
the communication poses a relatively small overhead to the calculation
cost, in our largest simulation the communication overhead $\aplt
10$\%. We have therewith demonstrated that large-scale structure
formation simulations are excellently suited for high-performance grid
computing.

The adopted setup would in principle have reduced our wall-clock time
by almost a factor two compared to running on a single supercomputer,
and our simulations in Tab.\,\ref{Tab:Simulation} have indeed
effectively benefitted from using the grid.  However, we have spend
more than a year preparing and optimizing the code, and acquiring and
scheduling the resources, none of which proved to be trivial.  On the
other hand, if we would not have opted for running on a grid it would
have proven extremely difficult to perform the simulations at all
since acquiring 1024 processors on a supercomputer for half-year via a
'normal' proposal would be challenging.  Even acquiring half the
resources on two supercomputers with the promise to switch computers
half-way the simulation would be difficult. Our strategy to run on a
grid enabled us to secure the required CPU time on both
supercomputers.  

During this project we all became very enthusiastic about the
computational grid as a high-performance resource, despite the time we
have spend with preparations.  The success of our grid is in part a
consequence of the realization that latency forms no bottleneck, even
if the computers are separated by half the planet. We cannot beat the
speed of light, but bandwidth is likely to improve with time, making
high-performance grid computing very attractive in the foreseeable
future. Much work, however, is needed in improving practical matters,
like scheduling issues, network acquisition and cooperation between
supercomputer centers, each of which are major bottlenecks.  We
seriously consider to perform our next run on a grid with many
supercomputers.

\begin{figure}[h]
   \psfig{figure=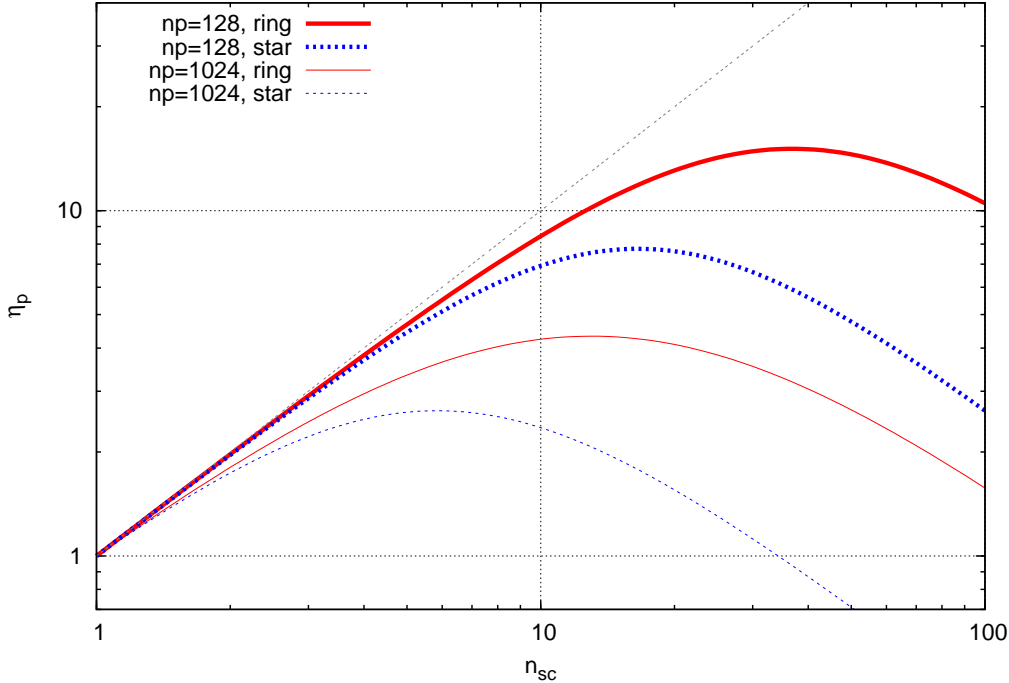,width=0.7\columnwidth,angle=-90} 
\caption[]{ 
   The speedup ($\eta_p$) for a grid of $n_{sc}$ supercomputers each
   with $p$ processors.  The speedup is defined as the wall-clock time
   of the grid computer as fraction of the wall-clock time of one
   single supercomputer, assuming that each computer has $p$
   processors.  The thick curves are calculated with $p=128$ the thin
   curves with $p=1024$.  The solid curves are calculated assuming the
   ring topology network, whereas for the dotted curves we adopted a
   star connected network. We adopted an average bandwidth of
   $b=200$\,Mbyte/s with a latency of $\nu t_{\rm lat}=0.828$\,s for
   the external (grid) communication and 10Gbyte/s for internal
   communication.  The diagonal thin dashed curve indicates the ideal
   scaling.
  } \label{fig:Prognosis} 
\end{figure}

Based on our measurements for each of the two supercomputers and the
intercontinental grid we constructed a performance model, which is
composed of two main components: the calculation ($t_{\rm cpu}$) and
the data transfer over the grid $t_{\rm comm} = \nu t_{\rm lat} +
S_{\rm comm}/b$, where $t_{\rm lat}$ is the network latency and $b$ is
the network throughput (see Eq.\,\ref{Eq:CommSize}). The parameter
$\nu$ is introduced to correct for the inefficiencies in our code
which require several transmissions ($\nu$) to the other computer.  In
Fig.\,\ref{fig:Prognosis} we present the results of the performance
model based on the characteristics of the adopted computers, network
and software environment for $N=2048^3$. The speedup is limited by the
bandwidth, whereas latency, for which we adopted $t_{\rm lat} =
0.138$s with $\nu =6$, poses no limiting factor to our
simulations. Reducing the total latency will hardly improve the
performance, but improving the throughput by a factor 10 would allow
us to use an order of magnitude more processors per supercomputer
while still acquiring acceptable speedup.  The two topologies adopted
in Fig.\,\ref{fig:Prognosis} are selected based on an ideal setup
where the supercomputers are interconnected via a ring topology.  A
sub-optimal star topology leads to congestion as the communication
tends to go through a single site.  With the optimal topology (solid
curves in Fig.\,\ref{fig:Prognosis}) running on a 10 supercomputers
with $\aplt 128$ processors each results in a speed-up of a factor of
$\sim 8$, whereas running on 100 supercomputers the speedup would be
only a factor of 10.


The best strategy for performing cold dark matter simulations using a
treePM code on a grid appears to be by acquiring $p \aplt 100$
processors on $n_{\rm sc} \sim 10$ supercomputers in a ring network
topology.  The main reason for adopting this strategy would be to be
able to acquire the required compute resources; it is considerably
easier to obtain $p \sim 100$ processors for an extended period ($\sim
1$~year) from a dozen supercomputer centers, than to acquire the
required CPU hours on a single supercomputer.  This strategy, however,
would require drastic changes in the co-scheduling policy of the
supercomputer centers.

With our cosmological cold dark matter simulation we make the dream of
Foster and Kesselman \cite{FosterKessleman} comes true, our
calculation benefits from having multiple widely separated computers
interconnected with a high-bandwidth network, effectively operating as
a single machine.

\section*{Acknowledgments}

We are grateful to Hans Blom, Maxine Brown, Andreas Burkert, Halden
Cohn, Jonathan Cole, Tom Defanti, Jan Eveleth, Katsuyuki Hasabe,
Douglas Heggie, Wouter Huisman, Piet Hut, Mary Inaba, Akira Kato,
Walter Lioen, Kees Neggers, Breanndan \'O Nu\'aill\'an, Hanno Pet,
Steven Rieder, Joaching Stadel, Huub Stoffers, Yoshino Takeshi, Thomas
Tam, Jin Tanaka, Peter Tavenier, Mark van de Sanden, Ronald van der
Pol Alan Verlo and Seiichi Yamamoto.
We are grateful to Ben Moore for offering us a bottle of {\em Pommery
1998 Cuv\'ee Louise Brut} for the first relevant scientific results
coming from this calculation.
This work was supported by NWO (grants \#643.200.503
and \#639.073.803), the NCF (project \#SH-095-08), QosCosGrid
(EU-FP6-IST-FET Contract \#033883), NSF IRNC, NAOJ, NOVA, LKBF and the
JSPS.
We thank the network facilities of SURFnet, Masafumi Ooe; IEEAF; WIDE;
Northwest Gigapop and the Global Lambda Integrated FAcility (GLIF)
GOLE of TransLight Cisco on National LambdaRail, TransLight,
StarLight, NetherLight, T-LEX, Pacific and Atlantic Wave.  
The calulations were performed on the supercomputers at Cray XT4 at
the Center for Computational Astrophysics of National Astronomical
Observatory of Japan and the national academic supercomputer center
SARA in Amsterdam, the Netherlands.

\section*{The authors}
\paragraph{Simon Portegies Zwart}
is professor in computational astrophysics at the Sterrewacht Leiden
in the Netherlands. He received his Ph.D. in astronomy at Utrecht
University.  
His principal scientific interests are high-performance
computational astrophysics and the ecology of dense stallar systems.\\
e-mail:{\tt spz@strw.leidenuniv.nl},\\
URL:{\tt http://www.strw.leidenuniv.nl/$\sim$spz/}

\paragraph{Tomoaki Ishiyama and Derek Groen}
Are graduate students in the research groeps of Makino and Portegies
Zwart, respectivly. 

\paragraph{Jun Makino}
is professor of astrophysics at the National Astronomical Observatory
of Japan. He recieved his PhD in astronomy at the University in
Tokyo. 
His research interests are stellar dynamics, large-scale
scientific simulation and high-performance computing.\\ 
e-mail:{\tt makino@cfca.jp},\\
URL:{\tt http://www.artcompsci.org/$\sim$makino}

\paragraph{Cees de Laat}
is associate professor of system and network engineering 
at the University of Amsterdam.  He recieved his PhD in computer
science at the University of Amsterdam.  
His research interests are
optical/switched Internet for data-transport in TeraScale eScience
applications.
e-mail:{\tt delaat@uva.nl},\\
URL:{\tt http://www.science.uva.nl/$\sim$delaat}

\paragraph{Steve McMillan}
is professor of physics at Drexel University in Philadelphia,
Pennsylvania, U.S.A.  He received his Ph.D. in astronomy from Harvard
University.  
His principal scientific interests are high-performance
computation and the astrophysics of star clusters and galactic nuclei.\\
e-mail:{\tt steve@physics.drexel.edu},\\
URL:{\tt http://www.physics.drexel.edu/$\sim$steve}

\paragraph{Kei Hiraki}
is professor in computer science at the University of Tokyo.  He
received his Ph.D. in phisics at the University of Tokyo. 
His research
interests are parallel and distributed system, high-performance
computing and networking.\\ 
e-mail:{\tt hiraki@is.s.u-tokyo.ac.nl},\\
URL:{\tt http://www-hiraki.is.s.u-tokyo.ac.jp}

\paragraph{Stefan Harfst}
is a postdoctoral fellow in the computational astrophysics group of
Portegies Zwart. He received his Ph.D. in astrophysics at the
University of Kiel, Germany.  
His research interests are stellar
dynamics and high-performance computing.\\
e-mail:{\tt harfst@strw.leidenuniv.nl}

\paragraph{Keigo Nitadori}
Is a postdoctoral fellow at RIKEN, Japan. He received his Ph.D. in
astrophysics at the University of Tokyo.\\ e-mail:{\tt keigo@riken.jp}

\paragraph{Paola Grosso}
is a senior researcher at the system and network engineering group at
the University of Amsterdam. She received her Ph.D. in Physics from
the University of Turin in Italy. 
Her research interests are
modeling, programming and virtualization of networks for eScience
applications.\\
e-mail:{\tt p.grosso@uva.nl},\\
URL:{\tt http://www.science.uva.nl/$\sim$grosso}

\end{document}